\documentclass[a4paper,11pt]{article}
\pdfoutput=1
\usepackage{geometry}
\usepackage[svgnames]{xcolor}
\usepackage[colorlinks=true,urlcolor=black,linkcolor=blue,citecolor=blue,bookmarks=false]{hyperref}
\usepackage{amsfonts,amsmath,amssymb}
\usepackage{graphicx}
\usepackage{dsfont}
\usepackage{graphicx}
\usepackage{bm}
\usepackage[font=small,labelfont=bf]{caption}
\renewcommand*{\thepage}{\footnotesize\arabic{page}}
\usepackage{color}
\usepackage{enumitem}
\usepackage{boldline}
\usepackage{changepage}
\usepackage{cite}
\usepackage{textcomp}
\usepackage[title]{appendix}
\usepackage{url}
\usepackage[T1]{fontenc}
\usepackage{authblk}
\usepackage{breakurl}
\usepackage{fancyhdr}
\usepackage[export]{adjustbox}
\usepackage[percent]{overpic}
\usepackage{mathpazo}
 \usepackage{multirow}
 \usepackage{float}
\usepackage{color}
\usepackage{soul}
\setstcolor{red}
\makeatletter
\def\SOUL@ulthickness{2pt}
\makeatletter

\usepackage{etoolbox}
\makeatletter
\patchcmd{\l@section}
  {\hfil}
  {\leaders\hbox{\normalfont$\m@th\mkern \@dotsep mu\hbox{.}\mkern \@dotsep mu$}\hfill}
\makeatother

\title{ \bf Spatiotemporal statistical features of velocity responses to traffic congestions in a local motorway network}
\author{Shanshan Wang \thanks{shanshan.wang@uni-due.de}, Michael Schreckenberg and Thomas Guhr}
\affil{\textit{Faculty of Physics, University of Duisburg--Essen,  Lotharstra\ss e 1, 47048 Duisburg, Germany}}

\date{\today}
 
\begin{document}
\maketitle

\noindent {\bf Abstract.} 
The causal connection between congestions and velocity changes at different locations induces various statistical features, which we identify and measure in detail. We carry out an empirical analysis of large-scale traffic data on a  local motorway network around the Breitscheid intersection in the North Rhine-Westphalia, Germany. We put forward a response function which measures the velocity change at a certain location versus time conditioned on a congestion at another location. We use a novel definition of the corresponding congestion indicator to ensure causality. We find that the  response of velocities to the congestion exhibits phase changes in time. A  negative response at smaller time lags transforms into positive one at  larger time lags, implying a certain traffic mechanism. The response decays as a power law with the distance. We also identify a scaling property leading to a collapse of the response functions on one curve.

\vspace{0.5cm}

\noindent{\bf Keywords\/}: response function, traffic congestion, power law, scale invariance
\vspace{1cm}

\noindent\rule{\textwidth}{1pt}
\vspace*{-1cm}
{\setlength{\parskip}{0pt plus 1pt} \tableofcontents}
\noindent\rule{\textwidth}{1pt}

\section{Introduction}
\label{section1}

The traffic flow on road networks~\cite{Hansen1959,Geurs2004,Saif2019,Meersman2017} consists of free flow and congested flow. According to the three-phases traffic theory~\cite{Kerner2012}, the congested flow further contains  synchronized flow and wide moving jams. Extensive studies on the dynamic behavior of traffic flow has been devoted to modeling and simulations in past decades~\cite{Nagel1992,Schadschneider1993,Lovaas1994,Schreckenberg1995,Hoogendoorn2001,Burstedde2001,Wong2002,Fellendorf2010,Treiber2013}. At present, neither models nor simulations fully capture the realistic traffic situations, which may be affected by commuting, weather, seasons, road construction, traffic accidents, big city events, etc. A huge amount of traffic data collected by the global positioning system tracking devices, inductive loop detectors, video recording devices, etc.~\cite{Leduc2008} is available, making the empirical studies possible~\cite{Kerner2002,Bertini2005,Schonhof2007,Kerner2012}. Along with theoretical studies, data-driven analyses to explore traffic flow dynamics~\cite{Treiber2013,Li2020}, traffic patterns~\cite{Chowdhury2000,Kerner2002,Kerner2012}, traffic congestion~\cite{Afrin2020,Krause2017}, traffic flow prediction~\cite{Lv2014,Abadi2014,Kan2019} and the resilience after traffic jams~\cite{Zhang2019,Tang2018} are called for and pose a variety of challenges.

Due to the non-stationary in the time series of traffic observables, including traffic flows and velocities, a traffic network can be viewed as a complex system, where traffic observables are correlated in various ways in time and space. Temporal correlation matrices together with the technique of $k$-means clustering have been used for identifying different quasi-stationary states in traffic systems~\cite{Wang2020,Wang2023a}. The states, manifesting themselves in correlation structures, carry certain traffic patterns, related to non-stationary. In contrast to a financial market~\cite{Wang2018}, the presence of spatial information~\cite{Gartzke2022} renders a traffic network more complicated. The correlations between time series measured at arbitrarily labeled locations or positions induce a topology which has to be mapped on the real topology, i.e. of the road map. This has led to the identification of collective and sub-collective traffic behavior in the motorway network of North Rhine-Westphalia~\cite{Wang2021,Wang2022}.

The propagation of effects among road sections via a traffic network takes time, inducing temporal shift of correlation structures. Non-synchronized time series from different road sections bring about cross-correlations with a time lag or lead. A recent study~\cite{Gabor2023} discloses a spectral transition in the symmetrized matrix of time-lagged correlations. Importantly, the spectral transition is associated with the duration of traffic congestion. In addition to correlations, response functions as a novel concept are introduced to explore the interaction of road sections with non-synchronized time series~\cite{Wang2023b}. They measure the time-dependent response of some observable, conditioned on events which are encoded in indicator functions. The response function has been used in financial markets to study how the trading price changes conditioned on a buy or a sell~\cite{Bouchaud2003,Wang2016a,Wang2016b,Wang2017,Benzaquen2017,Grimm2019,Henao2021}. It consists of a response variable and a triggered event. In traffic, the latter could be traffic congestion~\cite{Krause2017,Zhang2019,Tang2018}, traffic accidents~\cite{Saladie2020}, road construction~\cite{Fei2016}, the presence of trucks~\cite{Han2015}, etc.

Our previous study on the response of velocities to heavy congestion~\cite{Wang2023b} was conducted with five neighbouring sections on a motorway. The response function measures the average velocity changes versus time on a motorway section conditioned on the heavy congestion occurring on a different section at an earlier time. Despite the fact that a remarkable response with phase transitions shows up, it is difficult to build a causal relation between the velocity change and the heavy congestion, as congestion may occur simultaneously on other sections in addition to the section as the trigger. Furthermore, the degree of velocity changes on a section depends largely on its specific traffic environment, for instance, a section with bottleneck, a must-pass section for commuting, or a section on a bridge. In our previous study~\cite{Wang2023b}, we ignore the effect of traffic environment on the responses, as the considered sections are rather close to each other. When studying the responses among sections distributed on a motorway network in two dimensions, the environments for the sections may vary largely over the network, which has to be taken into account.

We extend the study of the considered motorway network from one to two dimensions. First, we introduce a conditional indicator, which rules out the possibility of synchronized congestion on multiple sections and guarantees the response only caused by the section as a trigger. Second, we employ an alternative definition of response functions, which removes the effect of random noise on velocity changes. Third, by considering many sections distributed on a two-dimensional motorway network, we are able to explore the spatial features of responses, in addition to the temporal features. 

This paper is organized as follows. In Sec.~\ref{sec2}, we provide some basic information and concepts for this study, including the considered local motorway network, the used traffic data, the method to aggregate velocities across multiple lanes, and the network distances. In Sec.~\ref{sec3}, we introduce the response functions for this study as well as the indicator functions that ensure causality. In Sec.~\ref{sec4}, we analyze our empirical results and find phase changes in time as well as power laws and scaling invariance in space. We conclude in Sec.~\ref{sec5}.

\section{Data description}
\label{sec2}
In Sec.~\ref{sec21}, we introduce a local motorway network considered in this study and the information of traffic data. In Sec.~\ref{sec22}, we describe the method of averaging velocities across multiple lanes on a motorway section. In Sec.~\ref{sec23}, we brief the network distance and its computation method.  

\subsection{The studied motorway network and traffic data}
\label{sec21}

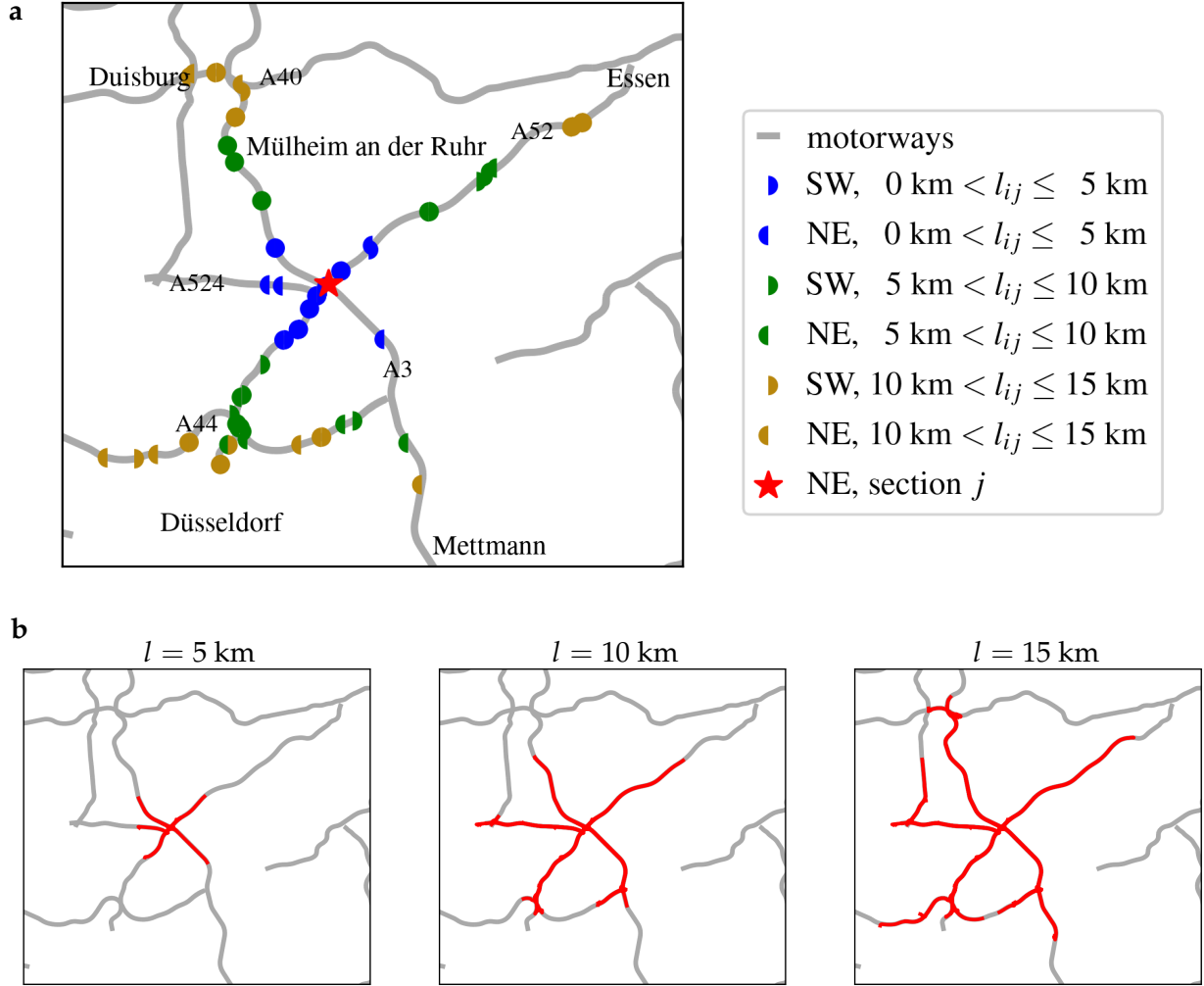
\begin{figure}[tb]
\begin{center}
\begin{overpic}[width=0.9\textwidth]{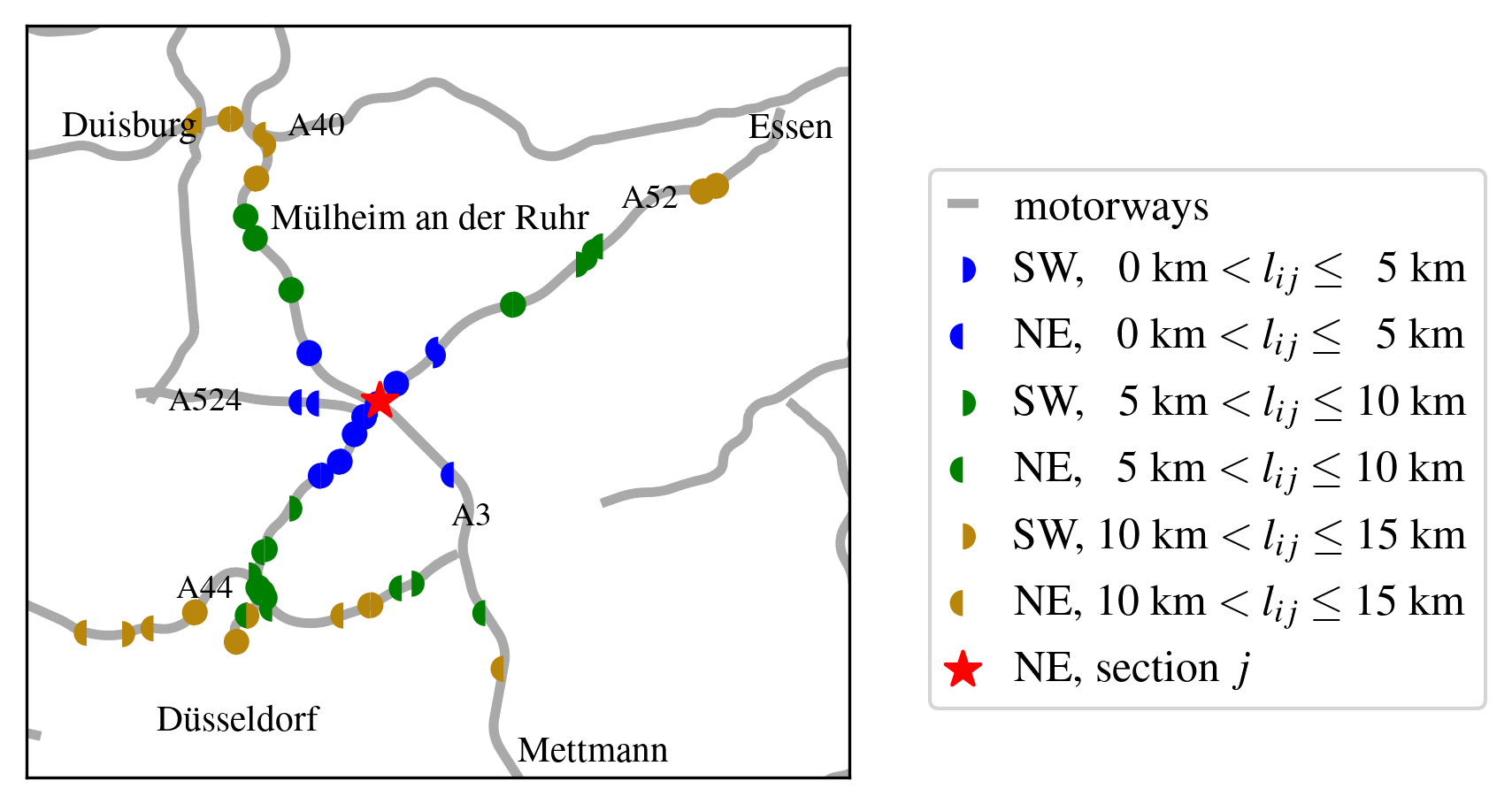}
\put(-3,50){\textbf a}
\end{overpic}\\ [1cm]
\begin{overpic}[width=0.95\textwidth]{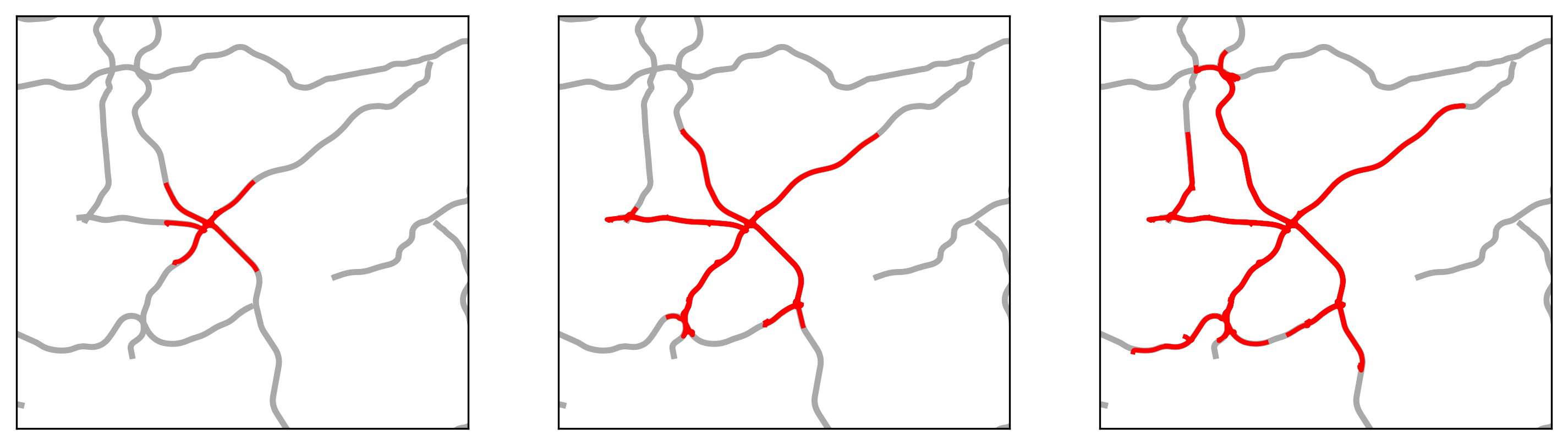}
\put(0,30){\textbf b}
\put(11,28){$l=5$ km}
\put(45,28){$l=10$ km}
\put(80,28){$l=15$ km}
\end{overpic}\\ [0.5cm]
\caption{{\textbf a} All together 68 sections $i$ toward the south-west (SW) or the north-east (NE) at different ranges of network distances $l_{ij}$ between them and one section $j$ in the centre of a local motorway network near the Breitscheid intersection in NRW, Germany. {\textbf b} Three examples of the covered areas of the motorway network within a given distance ranges $l$ (red line) from the central section $j$, where $l=5$, 10, and 15 km, respectively.}
\label{fig1}
\end{center}
\end{figure}

In this study, we focus on a local motorway network near Breitscheid in North Rhine-Westphalia (NRW), Germany, which is part of the large-scale NRW motorway network. The considered local network is mainly composed of motorway A52 connecting the densely populated cities of  D\"usseldorf and Essen, motorway A3 connecting the cities of Duisburg and Mettmann, motorway A40 connecting the cities of Duisburg and Essen, and motorway A524 connecting the city of Duisburg with other motorways, as displayed in Fig.~\ref{fig1}a. The intersection of motorways A3 and A52 is at Breitscheid and carries heavy traffic flow of commuters during rush hours on workdays. We select a section on motorway A52 as close to this intersection as possible to study the effect of its congestion on other sections nearby. As seen in Fig.~\ref{fig1}, this section toward the north-east (NE) is our section $j$ locating in the center of the network and playing the role of congestion. The other sections either toward north-east or toward south-west (SW) are the sections $i$ response to it. Within the network distance of 15 km from section $j$, we have 68 sections $i$ in total.

Our traffic data is accumulated with inductive loop detectors on the  motorway network. It includes the information on traffic flow and on velocity with a resolution of one minute for each lane on each motorway section. The data used in this study comprises 179 workdays selected during the period from Dec. 1, 2016 to Nov. 30, 2017. On each considered workday, our central section $j$ contains the velocity of at least one minute lower than or equal to 10 km/h from 5:00 to 22:59, which guarantees the presence of heavy congestion of section $j$ during this period. Moreover, the used data for each section from 5:00 to 22:59 on the 179 workdays has high quality with more than $96.7\%$ non-missing values. We fill the missing values in the data with the linear interpolation of neighboring, non-missing values~\cite{Wang2022}.

\subsection{Velocities on individual sections}
\label{sec22}

One motorway section has one or more lanes, leading to one or multiple velocities per minute. The later case requires aggregation of the velocity across multiple lanes, such that for one section there is one velocity per minute. For the velocity aggregation, we use the flow-weighted velocity here, which is different from the density-weighted velocity that we used in our previous studies~\cite{Wang2020,Wang2021,Gartzke2022,Wang2022,Wang2023a,Wang2023b}. 
The traffic flow is the number of vehicles passing through a road section per unit time, while the density is the number of vehicles passing through per unit distance. The former is a time-dependent observable obtained from data directly, while the latter is a space-dependent quantity that has to be worked out via flow and velocity. In contrast to density-weighted velocity, the flow-weighted velocity better reflects the velocity per individual vehicle.

\begin{figure}[tb]
\begin{center}
\begin{overpic}[width=0.95\textwidth]{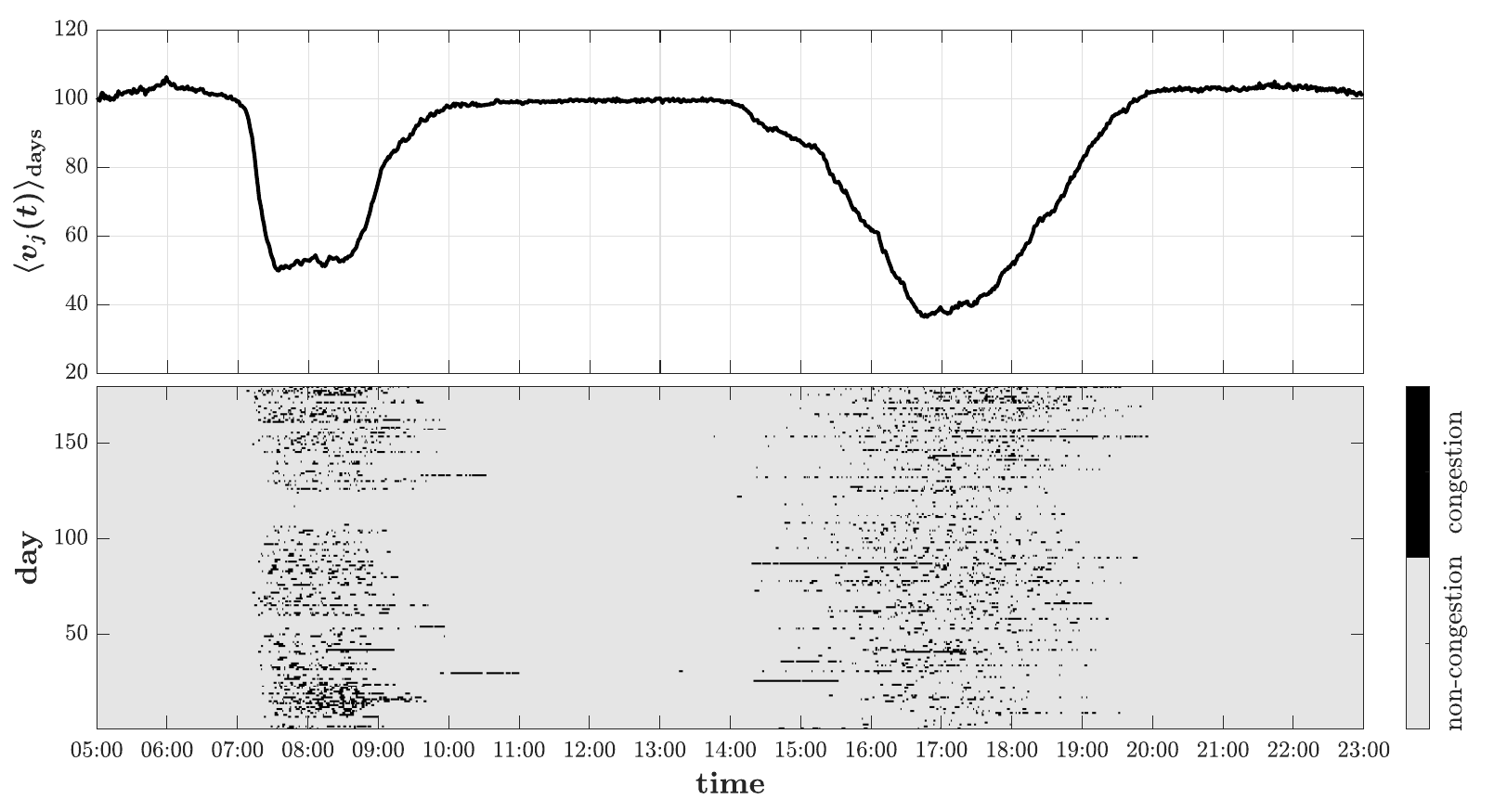}
\put(7,50.5){\textbf a}
\put(7,26){\textbf b}
\end{overpic}
\caption{{\textbf a} Velocity $\langle v_j(t)\rangle_\mathrm{days}$ of section $j$ averaged over 179 workdays containing congestion. {\textbf b} A matrix that reveals the distribution of congestion between 5:00 and 22:59 on the 179 workdays with the critical velocity $v_c=10$ km/h.}
\label{fig2}
\end{center}
\end{figure}

Let the traffic flow and velocity at time $t$ on lane $m$ of section $i$ be denoted $q_{i,m}(t)$ and $v_{i,m}(t)$, respectively. We define the flow-weighted velocity $v_i(t)$ for section $i$ across multiple lanes as the sum of the velocity times the flow on each lane divided by the total flow on this section, 
\begin{equation}
v_i(t)=\frac{\sum_m q_{i,m} (t)v_{i,m}(t)}{\sum_m q_{i,m}(t)}\ .
\label{eq2.2.1}
\end{equation}
Distinguishing car flows and truck flows, we further extend the above equation to
\begin{equation}
v_i(t)=\frac{\sum_m (q_{i,m}^{(c)} (t)v_{i,m}^{(c)}(t)+q_{i,m}^{(t)} (t)v_{i,m}^{(t)}(t))}{\sum_m (q_{i,m}^{(c)}(t)+q_{i,m}^{(t)}(t))}\ ,
\label{eq2.2.2}
\end{equation}
where the superscripts $(c)$ and $(t)$ indicate the quantities for cars and for trucks, respectively. For convenience, we refer to the flow-weighted velocity as velocity in the following. As an example, Fig.~\ref{fig2}a shows the time evolution of the velocity on the central section $j$ averaged over 179 workdays. Two valleys are visible for morning and afternoon rush hours, where the valley during afternoon rush hours is much deeper and wider. Setting the critical velocity at 10 km/h, the time period with lower velocity is the congested phase, otherwise is the non-congested phase, depicted in Fig.~\ref{fig2}b. As expected, many congestions, in particular short congestions, occur during rush hours, but less or even none exist during non-rush hours.

\subsection{Network distances}
\label{sec23}

A network distance is the distance of the shortest path on the network between two locations. As for our local motorway network, the locations at the ends of a path are the motorway sections. The path along the motorway network is composed of many short motorway pieces connecting two close locations. Each short piece is similar to a straight line and its distance is approximately a straight-line Euclidean or a geodetic distance. Therefore, a network distance of a path is the sum of distances of all short pieces along the path. In this way, we obtain network distances between any two sections with the help of the Java application Osmosis and the Python packages OSMnx and NetworkX. Exchanging the origin and the destination of two given sections $i$ and $j$, the distances in the unit of kilometers change very little. In view of this, the network distance from $i$ to $j$is equal to the network distance from $j$ to $i$, i.e. $l_{ij} = l_{ji}$.

Figure~\ref{fig1} visualizes the sections $j$ within different ranges of network distances to the central section $j$. The shortest path between two sections is a curve rather than a perfect straight-line. Thus, the sections $i$ within each distance range are not located in a ring or a circle centered around the central section $j$. As an example, Fig.~\ref{fig1} visualizes the sections within different distance ranges and the covered areas of the motorway network within a given distance range $l$ from the central section $j$.

 \section{Response functions}
 \label{sec3}

To study the response to congestion, we define an indicator function for a given critical velocity $v_c$ as
\begin{equation} 
\varepsilon_{j}(t)=\left\{
\begin{array}{rl}
1,&\mathrm{if~} v_i(t)<v_c \ ,\\
0,&\mathrm{if~} v_i(t)\geq v_c\ ,
\end{array}
\right. 
\label{eq3.1}
\end{equation}
where $\varepsilon_{i}(t)=1$ for congested traffic and $\varepsilon_{i}(t)=0$ for non-congested traffic. The three-phases theory~\cite{Kerner2012} provides a possible interpretation for congested traffic. In a local motorway network, simultaneous congestions may occur on multiple sections, obscuring the causality between congestion and velocity changes on different sections. To unambiguously disclose this causality, it is essential to capture the effect of congestion on one section without the interference of simultaneous congestions on others. We define an indicator of congestion on section $j$ under the condition that there are no congestions on other sections $k$. Furthermore, we do that with spatial resolution by only including network distances $l_{ij}$ smaller than a given distance threshold $l_\omega$. Hence we introduce
\begin{equation}
\omega_j(t|l_{\omega})=\varepsilon_j(t)\prod_{l_{kj}\leq l_\omega,~k\neq j}(1-\varepsilon_k(t)) \ .
\label{eq3.2}
\end{equation}
A simultaneous congestion on any section $k$ with $k\neq j$ implying $\omega_j(t|l_{\omega})=0$. In this way, it removes the contribution of congestion from multiple sections to the velocity change under consideration. When the congestion is absent in any section at time $t$ except for section $j$, the conditional indicator is $\omega_j(t|l_{\omega})=1$ and only the congestion on section $j$ contributes to the velocity change. 

A velocity change is also termed a velocity increment between times $t$ and $t+\tau$ on section $i$,
\begin{equation}
\Delta v_i(t,\tau)=v_i(t+\tau)-v_i(t) \ , 
\label{eq3.3}
\end{equation}
where $\tau$ is referred to as time lag. The velocity increment varies largely at different traffic environments. We define the response function of velocities to the conditional indicators given distance $l_\omega$ as,
\begin{equation}
R_{ij}(\tau|l_{\omega})=\langle \Delta v_i(t,\tau)\omega_j(t|l_{\omega})\rangle-\langle \Delta v_i(t,\tau)\rangle \langle\omega_j(t|l_{\omega})\rangle \ .
\label{eq3.4}
\end{equation}
The average $\langle \cdots \rangle$ is on the times $t$. The response function~\eqref{eq3.4} depends on the chosen critical velocity $v_c$. It measures, on average, how large the velocity on section $i$ relatively changes from time $t$ to $t+\tau$, if a congestion is only on section $j$ at time $t$. From a formal mathematical viewpoint, the response function is a time-lagged covariance. As one of the time series is an indicator, we prefer the term response functions. If $R_{ij}(\tau |l_{\omega})>0$, the two observables move in the same direction, i.e., the increase (or decrease) of the velocity change is accompanied by the increase (or decrease) of the conditional indicator. In contrast, the two quantities move in opposite directions when $R_{ij}(\tau |l_{\omega})<0$. The second term in Eq.~\eqref{eq3.4} is the unconnected part, hence the response vanishes if there is no mutual dependence between the congestions and the velocity change. It depends on the studied system if one finds it convenient to include this unconnected part.

The effect of congestion propagates both in time and in space via the neighbouring sections~\cite{Wang2023b}. A section geographically close to the congested section suffers more influences from the congestion than a section far away~\cite{Wang2023b}. As the network distance $l_{ij}$ between the impacted section $i$ and the congested section $j$ plays an important role in the congestion propagation, we incorporate the spatial information into the response function~\eqref{eq3.4}. To assess the spatial characteristics in a more general way, we average over all impacted sections $i$ in the region defined by $l_{ij}<l$,
\begin{equation}
\big \langle R_{ij}(\tau,l|l_{\omega})\big \rangle_{i}=\frac{\sum\limits_{i}R_{ij}(\tau |l_{\omega})\Theta (l-l_{ij})}{\sum\limits_{i}\Theta (l-l_{ij})} \ ,
\label{eq3.5}
\end{equation}
where the step function
\begin{equation}
\Theta (l-l_{ij})=\left\{
\begin{array}{rl}
1,& \mathrm{if~} l\geq l_{ij}\\
0,& \mathrm{if~} l<l_{ij}
\end{array}
\right. \ 
\label{eq3.6}
\end{equation}
extracts all sections $i$ that satisfy the condition of distances. The average in Eq.~\eqref{eq3.5} captures the response within the specified range, and washes out the noise in the individual response functions.

\section{Empirical results and discussion}
 \label{sec4}
 
To empirically work out the response functions, we first apply Eq.~\eqref{eq3.4} to the time series of each workday and then average the response values for each given $\tau$ over different workdays to obtain $R_{ij}(\tau|l_{\omega})$. Averaging $R_{ij}(\tau |l_{\omega})$ over different distance ranges $l$ by Eq.~\eqref{eq3.5} finally results in $\big \langle R_{ij}(\tau,l|l_{\omega})\big \rangle_{i}$. In the following, we first discuss the response behavior with respect to the time evolution in Sec.~\ref{sec41}. We then analyze the transitions of response phases in Sec.~\ref{sec42}. We also explore how the response changes with the increase of distance ranges in Sec.~\ref{sec43}. We further inspect the feature of the scale invariance in the response function in terms of distance ranges in Sec.~\ref{sec44}.   

\subsection{Time-dependent response behavior}
 \label{sec41}

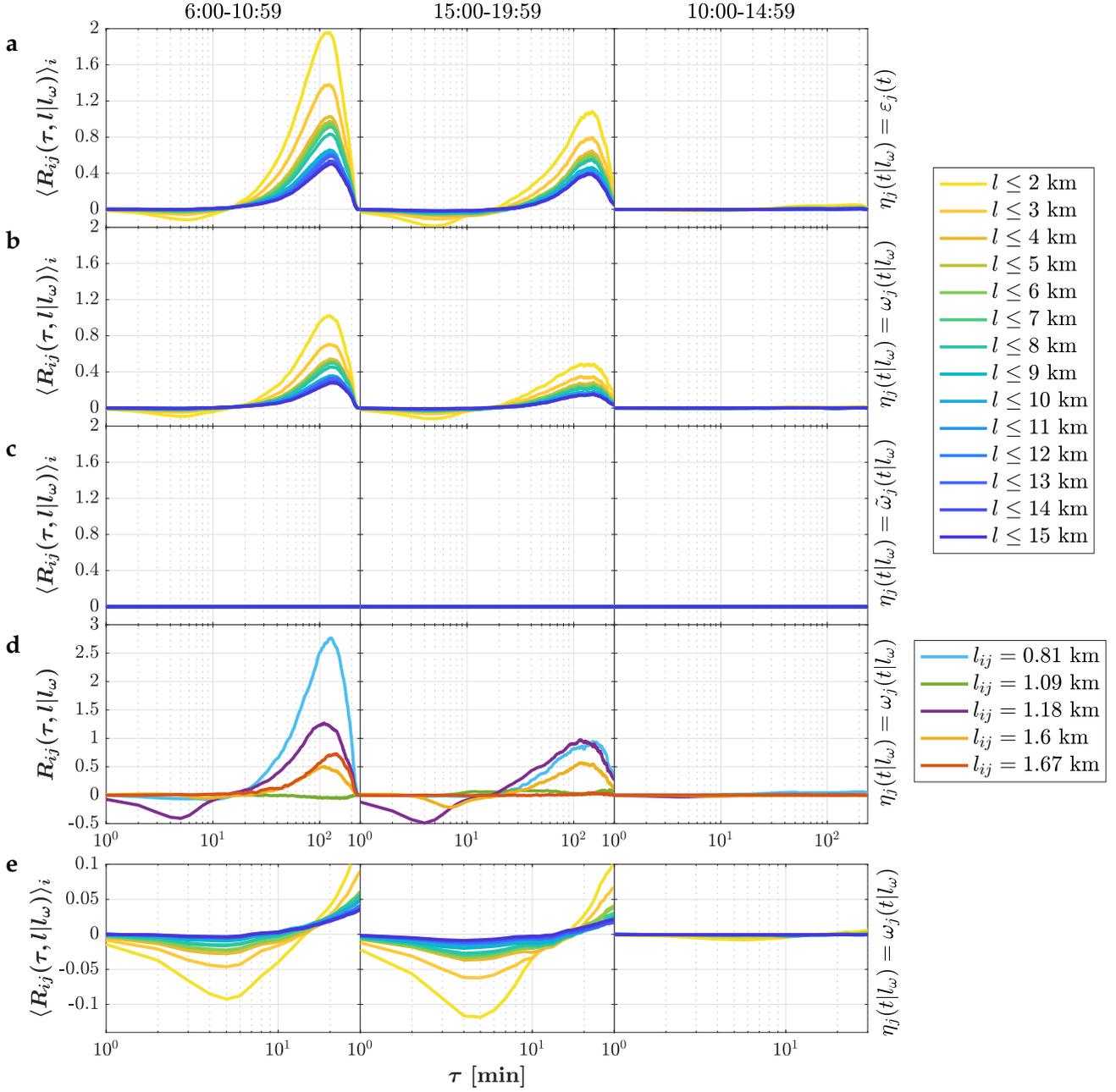
\begin{figure}[htbp]
\begin{center}
\begin{overpic}[width=\textwidth]{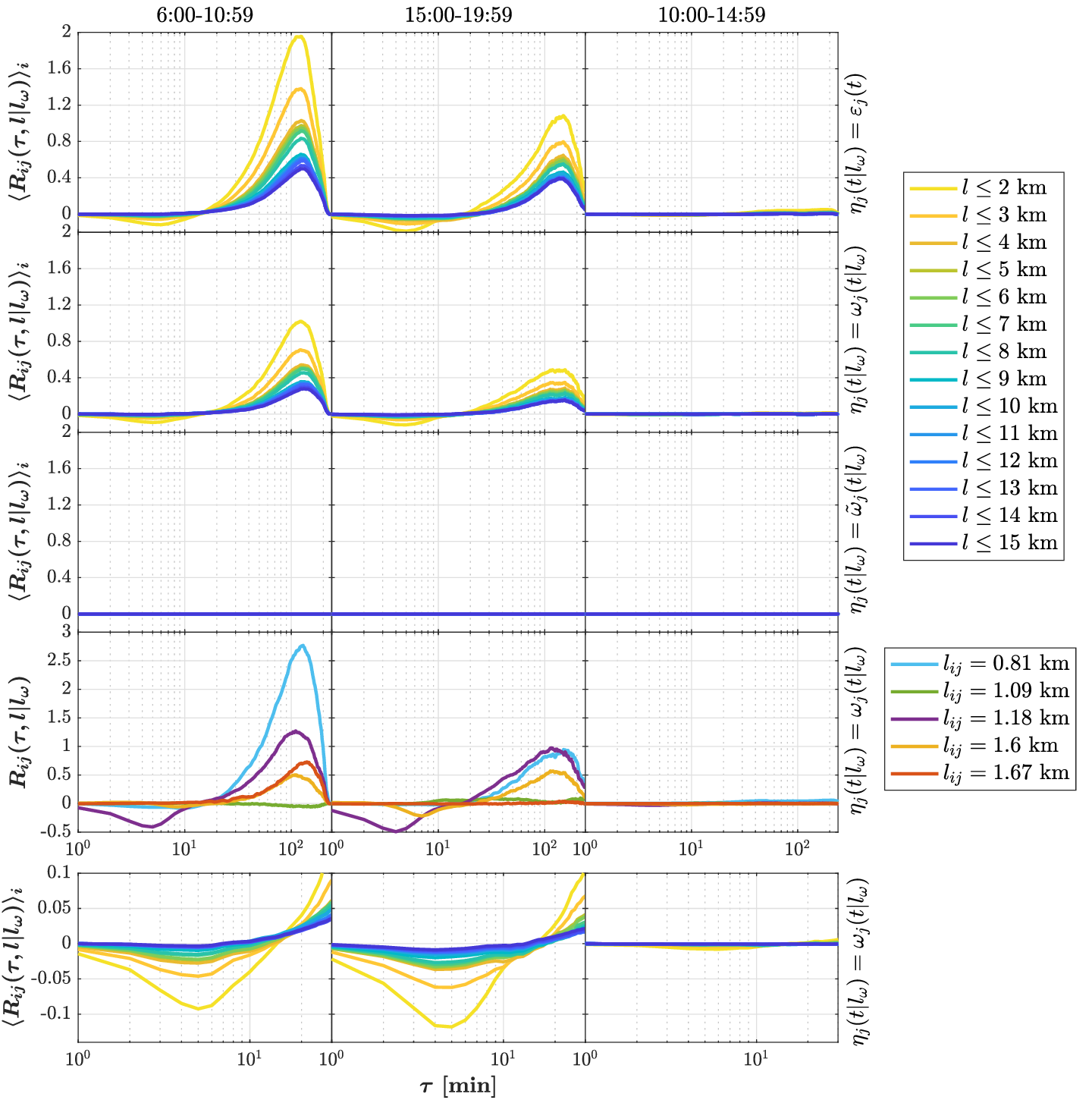}
\put(-2,95){\textbf a}
\put(-2,77){\textbf b}
\put(-2,58){\textbf c}
\put(-2,40){\textbf d}
\put(-2,20){\textbf e}
\end{overpic}
\caption{{\textbf a}, responses $\langle R_{ij}(\tau,l |l_{\omega})\rangle_i$ versus time lag $\tau$ with $\eta_j(t|l_\omega)=\varepsilon_j(t)$ in Eq.~\eqref{eq4.1.1} for congestions on section $j$ regardless of any simultaneous congestion on other sections $i$. {\textbf b}, responses $\langle R_{ij}(\tau,l |l_{\omega})\rangle_i$ versus time lag $\tau$ with $\eta_j(t|l_\omega)=\omega_j(t|l_\omega)$ for congestions only on section $j$. {\textbf c}, responses $\langle R_{ij}(\tau,l |l_{\omega})\rangle_i$ versus time lag $\tau$ with $\eta_j(t|l_\omega)=\tilde{\omega}_j(t|l_\omega)$ for congestions occurring simultaneously on all sections. {\textbf d}, responses $\langle R_{ij}(\tau,l |l_{\omega})\rangle_i$ of individual section pairs versus time lag $\tau$ with $\eta_j(t|l_\omega)=\omega_j(t|l_\omega)$. {\textbf e}, the enlargement of negative responses $\langle R_{ij}(\tau,l |l_{\omega})\rangle_i$ versus time lag $\tau$ with $\eta_j(t|l_\omega)=\omega_j(t|l_\omega)$. Each line indicates the response within each distance range $l$. We compare the responses during morning rush hours 6:00--10:59 (left column), afternoon noon rush hours 15:00--19:59 (middle column), and non-rush hours 10:00-14:59 (right column). }
\label{fig3}
\end{center}
\end{figure}

According to Fig.~\ref{fig2}, we select three typical time periods, i.e. morning rush hours from 6:00 to 10:59, afternoon rush hours from 15:00 to 19:59, and non-rush hours from 10:00 to 14:59. Each time period contains 300 minutes with a time step of 1 minute. Considering a motorway network centered around section $j$ within the largest reachable network distance $l_\omega=15$ km for conditional indicator $\omega_j(t)$, we work out the averaged responses of velocities on sections $i$ to the congestion on section $j$, shown in Fig.~\ref{fig3}b, within different distance ranges $l$ running from 2 km to 15 km at an increment of 1 km. Here the range within $l=1$ km only contains one section $i$ and the averaging of results is unable to eliminate the individuality carried by section $i$ paired with section $j$. We therefore ignore this case. A strength difference in individuality is visible in Fig.~\ref{fig3}d. In spite of it, the basic characteristics of response curves are similar. 

Figure~\ref{fig3}b depicts the overall characteristics of responses and correspondingly Fig.~\ref{fig3}e zooms in the negative responses at small $\tau$. Within each $l$, the averaged response $\langle R_{ij}(\tau,l |l_{\omega})\rangle_i$ depending on time lag $\tau$ drops down to be negative and then raises up to be positive. The negative value persists for more than 10 minutes until the positive value shows up. Such behavior emerges from the both morning and afternoon rush hours. In contrast, the response is too weak to be observed during non-rush hours. Usually the congested phases dominate most of time during rush hours, while non-congested phases are prevalent most of time during non-rush hours. The comparison between rush and non-rush hours turns out that the presence of remarkable responses is stimulated by the congested phase rather than the non-congested phase. Essentially, the response function is a covariance function which reveals the collective motion of two quantities, e.g. $\Delta v_i(t,\tau)$ and $\omega_j(t)$ in our study. The case of $\omega_j(t)=0$ is complicated. It corresponds to the non-congested phase on section $j$ and the congested phase on both section $j$ and any section $i$. Therefore the contribution to response with $\omega_j(t)=0$ is difficult to be distinguished. Differently, the case of $\omega_j(t)=1$ only corresponds to the congested phase on section $j$ accompanied with non-congested phases on all sections $i$. The resulting response is causally related to the congestion on section $j$ to some extent. For the negative response at small $\tau$, when the binary conditional indicator $\omega_j(t)=1$, the velocity changes $\Delta v_i(t,\tau)$ relative to the average velocity change caused by noise information become negative, implying the velocity on section $i$ decreases due to the congestion on section $j$. On the other hand, for the positive response at large $\tau$, the $\Delta v_i(t,\tau)$ relative to its average become positive when $\omega_j(t)=1$, suggesting the velocity increases on section $i$ conditioned on the congestion on section $j$.   

For comparison, we also work out the responses with regard to different types of indicators, given in a uniform formula by
\begin{equation}
R_{ij}(\tau|l_{\omega})=\langle \Delta v_i(t,\tau)\eta_j(t|l_{\omega})\rangle-\langle \Delta v_i(t,\tau)\rangle \langle\eta_j(t|l_{\omega})\rangle \ .
\label{eq4.1.1}
\end{equation}
When the indicator $\eta_j(t|l_{\omega})=\omega_j(t|l_{\omega})$, we arrive at the response function~\eqref{eq3.4} with respect to the congestions only occurs on section $j$. When $\eta_j(t|l_{\omega})=\varepsilon_j(t)$, there are responses to the congestion on section $j$ regardless of the simultaneous congestions on other sections $i$, see Fig.~\ref{fig3}a. Obviously, this response is stronger than the response to the congestion only on section $j$ comparing Figs.~\ref{fig3}a and b, since the former contains a part of responses to other sections $i$. In other words, the latter exactly excludes the response components caused by other sections $i$ apart from section $j$, so as to preserve the causality between each section pair. For an extreme scenario, every section in the considered local motorway network is congested. In this scenario, the conditional indicator is defined as
\begin{equation}
\tilde{\omega}_j(t|l_{\omega})=\prod_{l_{kj}\leq l_\omega}\varepsilon_k(t) \ .
\label{eq4.1.2}
\end{equation}
Setting $\eta_j(t|l_{\omega})=\tilde{\omega}_j(t|l_{\omega})$ in Eq.~\eqref{eq4.1.1} yields the response to the congestion on every section. It is, however, zero response for each $\tau$ shown in Fig.~\ref{fig3}c, as the aforementioned scenario is rather impossible to reach unless the whole local motorway network is broken down. From our empirical data, we have not met this scenario so far.  

\subsection{Transitions of response phases}
 \label{sec42}
 
 \begin{figure}[b!]
\begin{center}
\includegraphics[width=\textwidth]{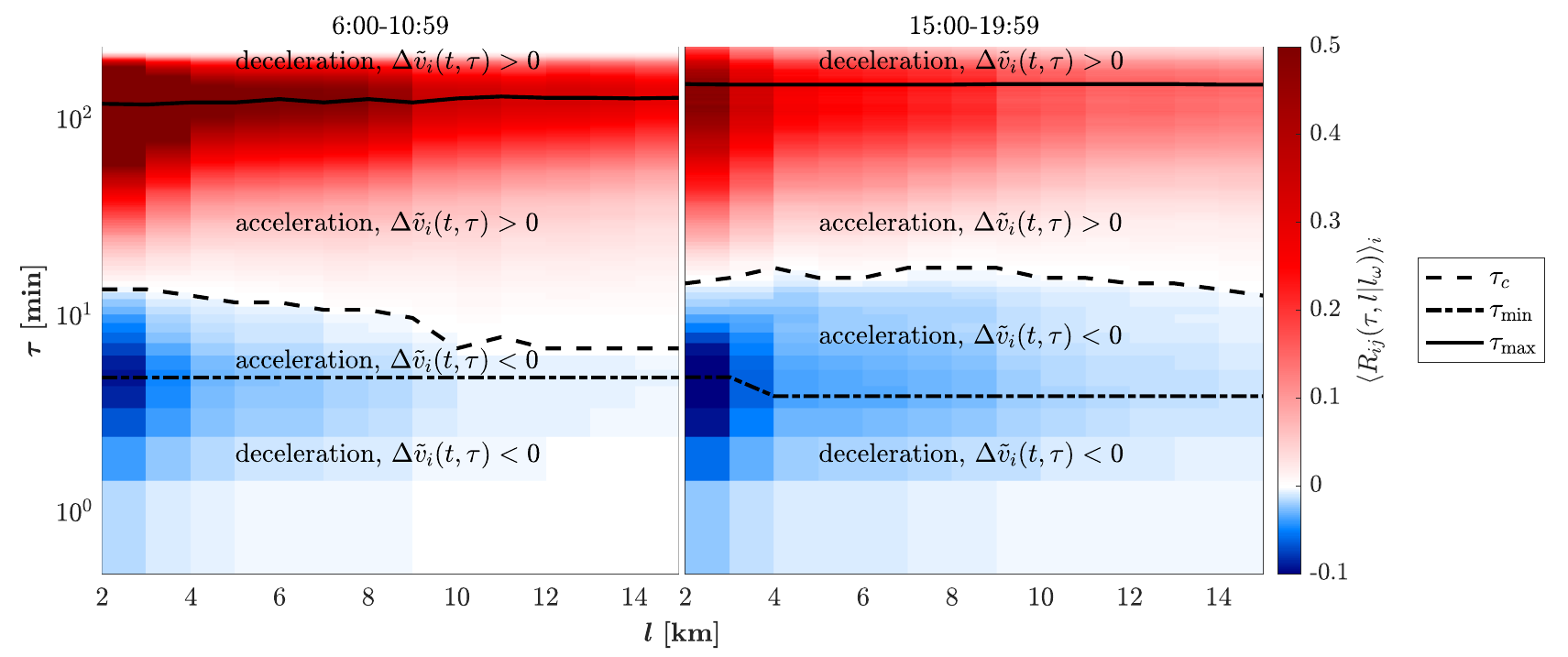}
\caption{Phase portraits of responses $\langle R_{ij}(\tau,l |l_{\omega})\rangle_{i}$ for morning rush hours 6:00--10:59 and afternoon rush hours 15:00--19:59, where the critical point $\tau_c$ and the points $\tau_\mathrm{min}$ and $\tau_\mathrm{max}$ separate the responses across the time lag $\tau$ and the distance range $l$ into four regions. In each region, we indicate the vehicle acceleration or deceleration as well as the velocity change relative to its average $\Delta \tilde{v}_i(t,\tau)$.}
\label{fig4}
\end{center}
\end{figure}

To explore the traffic dynamics from the perspective of velocity responses, we analyze different regimes. First, negative and positive responses are separated by the critical point $\tau_c$  ($0<\tau_c<30$ min), at which the response vanishes, 
\begin{equation}
\langle R_{ij}(\tau,l |l_{\omega})\rangle_i \Big |_{\tau=\tau_c}=0  \ .
 \label{sec4.2.2}
\end{equation}
In our previous study~\cite{Wang2023b}, we refer to the response occurring before $\tau_c$ as transient response (or response phase 1) and after $\tau_c$ as long-term response (or response phase 2). The phase 1 (phase 2) with the negative (positive) response reveals the lowering (raising) of the velocity on section $i$ caused by the congestion on section $j$. Furthermore, the response has a minimum at $\tau_\mathrm{min}$ ($0<\tau_\mathrm{min}<30$ min) and a maximum at $\tau_\mathrm{max}$ ($30$ min $<\tau_\mathrm{max}<240$ min), where its derivative vanishes,
\begin{equation}
\frac{\partial}{\partial \tau}\langle R_{ij}(\tau,l |l_{\omega})\rangle_i \Big |_{\tau=\tau_\mathrm{min}~\mathrm{or}~\tau_\mathrm{max}}=0 \ . 
 \label{sec4.2.3}
\end{equation}
The extremal points may be viewed as indicating transitions, reflecting competitions between the vehicle deceleration and acceleration on the impacted sections $i$. The three critical points separate the response in time and space into four regions, yielding a phase portrait for each rush hours, as depicted in Fig.~\ref{fig4}. For $0<\tau\leq \tau_\mathrm{min}$, the congestion causes a high possibility of vehicle deceleration, resulting in the decrease of the velocity on section $i$. Vehicles decelerate to a minimal value at around 5 or 4 minutes for different distance ranges. In comparison to the initial velocity, the velocity on section $i$ changes negatively. The magnitude of velocity changes decays with distance ranges. For $\tau_\mathrm{min}<\tau\leq\tau_c$, with the congestion relief, vehicle acceleration occupies the most of time, leading to the increase of the velocity from a negative value to the initial value. Roughly speaking, the larger the distance ranges, and the more quickly the vehicles recover to their initial velocities. The persistent acceleration during $\tau_c<\tau\leq\tau_\mathrm{max}$ further drives the velocity to a positive value. The vehicle acceleration for a long time attracts more traffic flow, which further reverses the change of velocity and leads to a reduction in velocity during $\tau_\mathrm{max}<\tau\leq240$ min.

\subsection{Power laws}
 \label{sec43}

\begin{figure}[tb!]
\begin{center}
\hspace*{0.5cm}
\begin{overpic}[width=\textwidth]{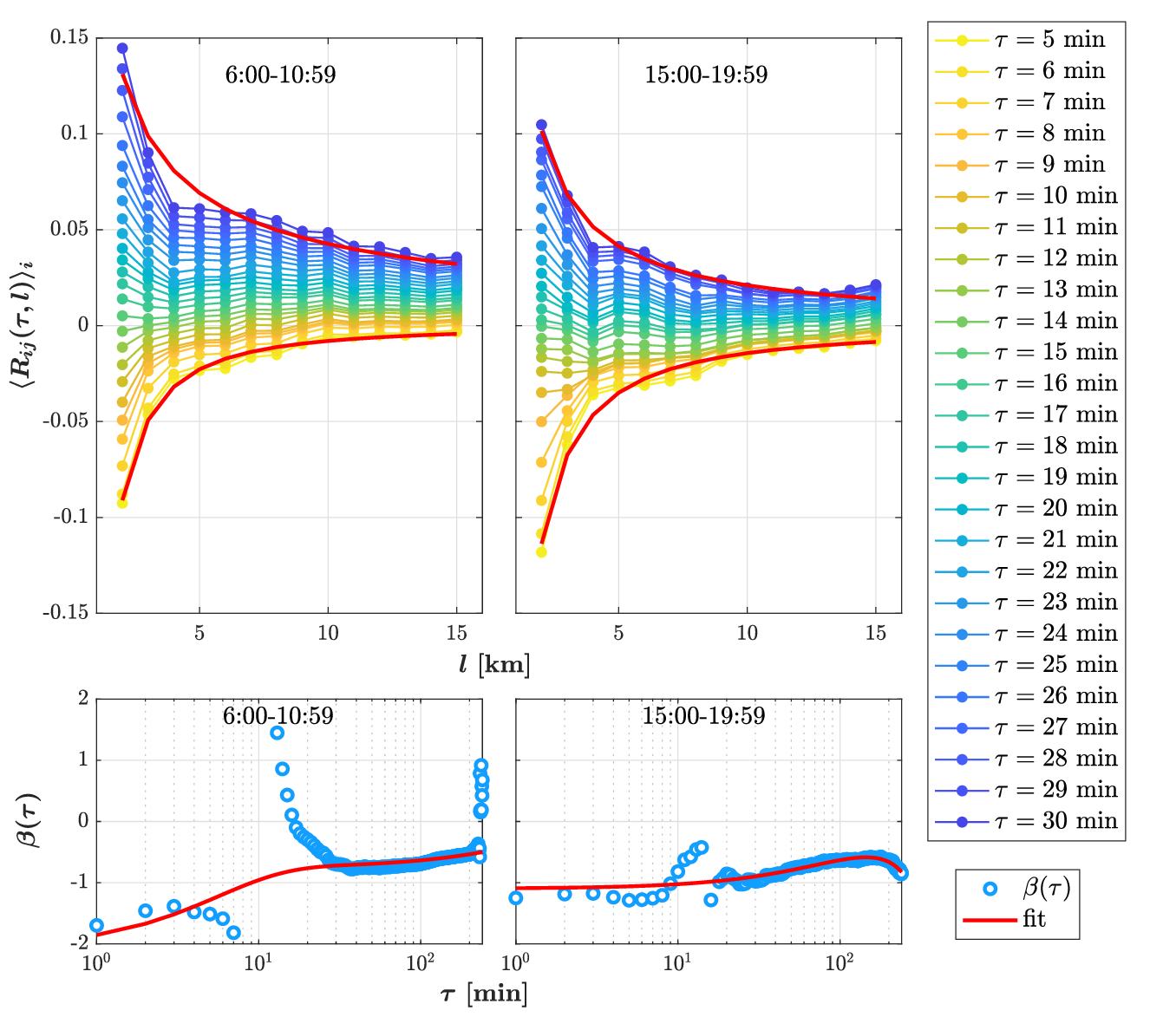}
\put(0,82){\textbf a}
\put(0,25){\textbf b}
\end{overpic}
\caption{{\textbf a}, the dependence of responses $\langle R_{ij}(\tau,l |l_{\omega})\rangle_{i}$ on the distance range $l$ at each given $\tau$ from 5 min to 30 min for morning and afternoon rush hours, where the dependences given $\tau=5$ min and $\tau=30$ min are fitted by power law~\eqref{eq4.3.1} with red lines. {\textbf b}, exponent $\beta (\tau |l_{\omega})$ resulting from power law~\eqref{eq4.3.1} versus $\tau$, where the values of $\beta(\tau |l_{\omega})$ fitted by exponential function~\eqref{eq4.4.6} to eliminate the anomalous behavior around the critical point $\tau_c$. Here $\tau_c\approx 9.5$ min for morning rush hours and $\tau_c\approx 15.5$ min for afternoon rush hours.}
\label{fig5}
\end{center}
\end{figure}

Figures~\ref{fig3}b,e and \ref{fig4} reveal that the response is not only time-dependent but also distance-dependent. Fixing a specific time lag $\tau$, the dependence of responses on the distance range $l$, as shown in Fig.~\ref{fig5}a, behave as a power law
\begin{equation}
\big \langle R_{ij}(\tau,l |l_{\omega})\big \rangle_{i}=\alpha(\tau |l_{\omega}) l^{\beta(\tau |l_{\omega})} \ ,
\label{eq4.3.1}
\end{equation}
where $\alpha(\tau |l_{\omega})$ is the $l$-independent part and $\beta(\tau |l_{\omega})$ the exponent. Both depend on $\tau$ and $l_\omega$. We determine them by fitting to the empirical result, as shown in Fig.~\ref{fig5}a. Around the critical point $\tau_c$, such a fit is not possible with statistical significance. This region has to be excluded. The results for $\beta(\tau |l_{\omega})$ are shown in Figs.~\ref{fig5}b and c for morning and afternoon rush hours during workdays. As seen, the exponent $\beta(\tau |l_{\omega})$ depends on the time lag $\tau$ considered. Importantly, there is a jump occurring in the region around $\tau_c$.

\subsection{Scale invariance}
 \label{sec44}

\begin{figure}[t]
\begin{center}
\includegraphics[width=\textwidth]{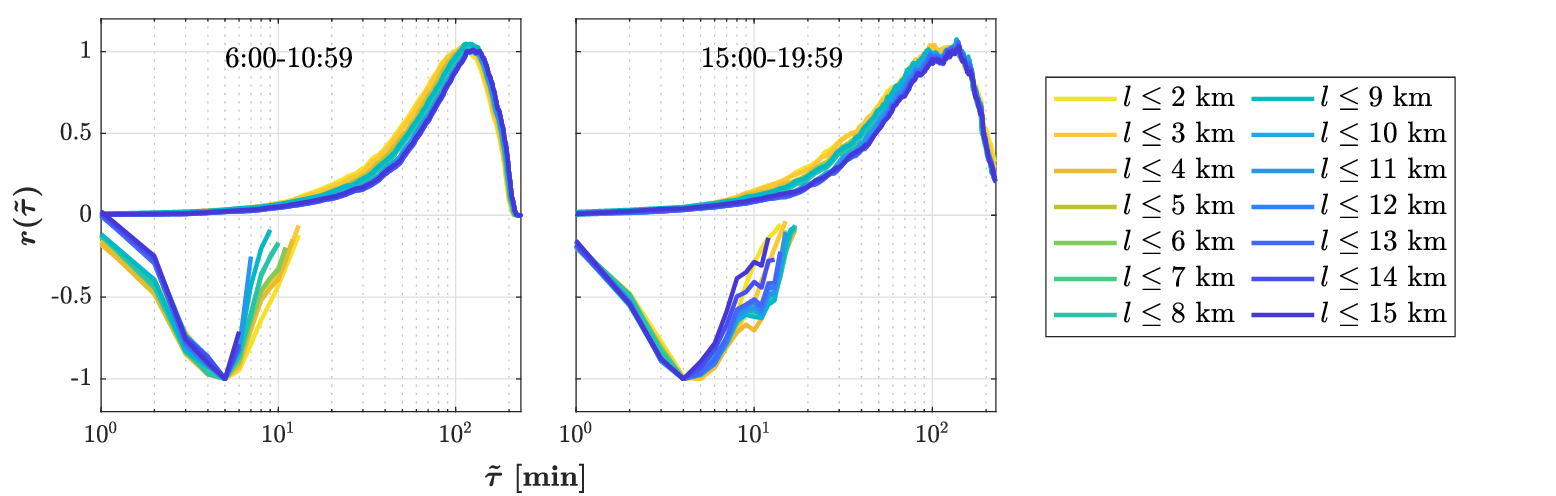}
\caption{Rescaled responses $r(\tilde{\tau})$ versus time lags $\tilde{\tau}$ within different distance ranges $l$ during morning and afternoon rush hours, where $\tilde{\tau}=\tau$ and $r(\tilde{\tau})< 0$ for $\tau<\tau_c$, and  $\tilde{\tau}=\tau-\tau_c$ and $r(\tilde{\tau})\geq  0$ for $\tau\geq \tau_c$.}
\label{fig6}
\end{center}
\end{figure}

Guided by our naked eyes, we phenomenologically describe the collapses of curves in Fig.~\ref{fig3}b by shifting horizontally and stretching vertically. To obtain well curve collapses, the responses before and after the critical point $\tau_c$ are rescaled by different methods,
\begin{equation}
r(\tilde{\tau})=\left\{
\begin{array}{ll}
\frac{\langle R_{ij}(\tau,l|l_{\omega})\rangle_i}{|\min(\langle R_{ij}(\tau,l|l_{\omega})\rangle_i)|} \ , \quad & \mathrm{with}~ \tilde{\tau}=\tau, ~\mathrm{if}~\tau<\tau_c \ , \\ [0.5cm]
\frac{\langle R_{ij}(\tau-\tau_c,l|l_{\omega})\rangle_i}{|\max(\langle R_{ij}(\tau,l|l_{\omega})\rangle_i)|} \ , \quad & \mathrm{with}~ \tilde{\tau}=\tau-\tau_c, ~\mathrm{if}~\tau\geq \tau_c \ .
\end{array}
\right. 
\label{eq4.4.1}
\end{equation}
For time lags $\tau<\tau_c$, the response is rescaled only by dividing the magnitude of the minimal response. After $\tau_c$, the response is not only shifted left by $\tau_c$, but also divided by the magnitude of the maximal response. The rescaled responses $r(\tilde{\tau})$ versus the time lag $\tilde{\tau}$ in Fig.~\ref{fig6} show that all curves are very close to each other and roughly collapse to a single curve. This phenomenon indicates a potential presence of scaling invariance. Furthermore, the difference in rescaling methods before and after $\tau_c$ suggest distinguishable traffic dynamics for different response phases.

To validate and refine our findings, we explore the behavior of scaling invariance employing the power law~\eqref{eq4.3.1}. It is known~\cite{Newman2005} that the only solution of the scaling-invariant criterion is a power law. We sketch the reasoning for the response function $\big \langle R_{ij}(\tau,l |l_{\omega})\big \rangle_{i}$ in Appendix~\ref{appa}. Assuming that the function $\big \langle R_{ij}(\tau,l |l_{\omega})\big \rangle_{i}$ in terms of distance ranges $l$ is invariant under all rescalings, we have~\cite{Newman2005,Sornette2009}
\begin{equation}
\big \langle R_{ij}(\tau,l |l_{\omega})\big \rangle_{i}=\mu(\lambda,\tau|l_{\omega}) \big \langle R_{ij}(\tau,\lambda l|l_{\omega})\big \rangle_{i} \ ,
\label{eq4.4.2}
\end{equation}
where $\lambda$ is a scaling factor. According to Eqs.~\eqref{eq4.3.1} and \eqref{eq4.4.2}, $\mu(\lambda,\tau|l_{\omega})$ is a function in terms of $\lambda$ and $\tau$,
\begin{equation}
\mu(\lambda,\tau|l_{\omega})=\lambda^{-\beta(\tau |l_{\omega})} \ .
\label{eq4.4.3}
\end{equation}
We reformulate Eq.~\eqref{eq4.4.2} as
\begin{equation}
\big \langle R_{ij}(\tau,l |l_{\omega})\big \rangle_{i}=\lambda^{-\beta(\tau |l_{\omega})} \big \langle R_{ij}(\tau,\lambda l |l_{\omega})\big \rangle_{i}\ .
\label{eq4.4.3}
\end{equation}
Setting $l=1$ and $\lambda=l$ in the above equation yields
\begin{equation}
\big \langle R_{ij}(\tau,1 |l_{\omega} )\big \rangle_{i}=l^{-\beta(\tau |l_{\omega})} \big \langle R_{ij}(\tau,l |l_{\omega})\big \rangle_{i} \ ,
\label{eq4.4.4}
\end{equation}
which means for a given $\tau$, the responses for different distance ranges $l$ are rescaled to the response within $l=1$ by multiplying $l^{-\beta(\tau |l_{\omega})}$. In other words, at a given $\tau$, the points of responses for different distance ranges $l$ overlap with each other. For different $\tau$, the connection of all overlapping points performs a collapse curve. Therefore, if the scaling invariance exists, the time-dependent curves of responses rescaled by multiplying $l^{-\beta(\tau |l_{\omega})}$ should collapse to a single curve. 

\begin{figure}[t!]
\begin{center}
\begin{overpic}[width=\textwidth]{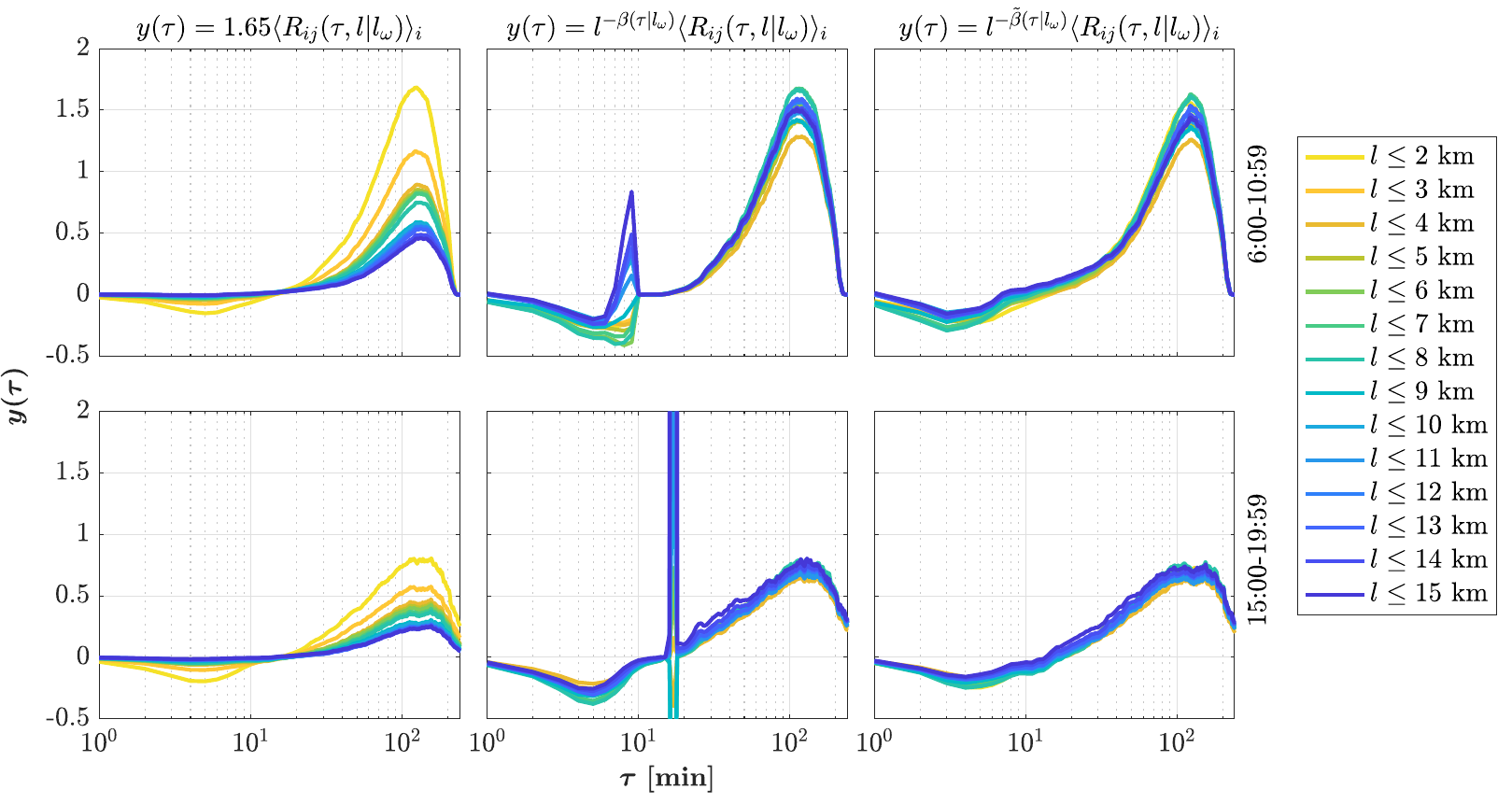}
\put(18,53.5){\textbf a}
\put(45,53.5){\textbf b}
\put(70,53.5){\textbf c}
\end{overpic}
\caption{Time evolution of responses $\langle R_{ij}(\tau,l |l_{\omega})\rangle_{i}$ within different distance ranges $l$ rescaled separately by multiplying a constant 1.65 ({\textbf a}), $l^{-\beta(\tau |l_{\omega})}$ ({\textbf b}), and $l^{-\tilde{\beta}(\tau |l_{\omega})}$ ({\textbf c}) during morning and afternoon rush hours.}
\label{fig7}
\end{center}
\end{figure}

Figure~\ref{fig7}b displays the empirical results of $l^{-\beta(\tau |l_{\omega})} \big \langle R_{ij}(\tau,l |l_{\omega})\big \rangle_{i}$ during morning and afternoon rush hours. Deviating from the critical point $\tau_c$, all curves within different distance ranges $l$ basically overlap with each other. As the region around $\tau_c$ does not allow a power-law analysis, we eliminate the effects from the critical point $\tau_c$ by fitting $\beta(\tau |l_{\omega})$ to an exponential function 
\begin{equation}
\beta(\tau |l_{\omega}) = a\exp(b\tau)+c\exp(d\tau)\ ,
\label{eq4.4.6}
\end{equation}
where $a$, $b$, $c$ and $d$ are fit parameters. The exponential function~\eqref{eq4.4.6} well describes the dependence of $\beta(\tau |l_{\omega})$ on $\tau$, as displayed in Fig.~\ref{fig5}c, and in particular fills suitable values to substitute for the distorted $\beta(\tau |l_{\omega})$ around $\tau_c$. For distinguishing, we refer to the fitted $\beta(\tau |l_{\omega})$ as $\tilde{\beta}(\tau |l_{\omega})$. With $\tilde{\beta}(\tau |l_{\omega})$, the rescaled responses almost collapse to the same curve regardless of the critical point $\tau_c$, as shown in Fig.~\ref{fig7}c. The overlap of all curves during afternoon rush hours looks much better than during afternoon rush hours. One possible reason owes to fitting errors either in the power law~\eqref{eq4.3.1} or in the exponential function~\eqref{eq4.4.6}. However, the most possible reason lies in the proportion of congestions during each rush hours. A higher proportion of congestions leads to the better statistic for responses, and further to the better collapses of rescaled responses. In Fig.~\ref{fig2}, a higher proportion of congestion exactly occupies the afternoon rush hours than the morning rush hours, corresponding to the better curve collapses for afternoon rush hours. Our empirical results, therefore, corroborate the assumption of scale invariance in the response function $\big \langle R_{ij}(\tau,l |l_{\omega})\big \rangle_{i}$ in terms of $l$ give each $\tau$. The values of exponent $\beta(\tau |l_{\omega} )$  before and after $\tau_c$ (see Sec.~\ref{sec43}) differ the scaling behavior for the response phases separated by $\tau_c$.

\section{Conclusions}
 \label{sec5}
 
To study the causality between the congestion and velocity changes, we introduced a new response function with a conditional indicator. The conditional indicator rules out the synchronization of congestion occurring on multiple motorway sections. The response function quantifies the causal connection between the impacted sections and the congested section. From a formal mathematical viewpoint, it is a (time-lagged) covariance. When the two quantities move towards the same direction, a positive response shows up and the velocity increases due to the congestion. Conversely, a negative response appears and the velocity decreases compared with the initial velocity. We found a phase change from negative responses at small time lags to positive responses at large time lags, separated by the critical point $\tau_c$ at which the response vanishes. The points $\tau_\mathrm{min}$ and $\tau_\mathrm{max}$ correspond to the minimal and the maximal response, respectively, where the minimal responses occur at around the time lag of 4 or 5 minutes. These points distinguish the vehicle deceleration from the vehicle acceleration. The latter leads to a velocity change relative to its average recovering from a negative value to a positive one. Therefore the acceleration prompts the change of response phases distinguished by the critical point $\tau_c$. The three points separate the response phases into four regions with different traffic dynamics. 

Furthermore, we also found the distance-dependent response at a fixed time lag $\tau$ decays as a power law in terms of the distance ranges within which the responses are averaged. We notice that a power law does not necessarily imply heavy tails, which depend on the exponent. Here, we focused on the scale invariance in response curves which we confirmed empirically.

\section*{Acknowledgments}

We are grateful to Sebastian Gartzke for fruitful discussions. We thank Strassen.NRW for providing the empirical traffic data. 

\section*{Author contributions}
T.G. and M.S. proposed the research. S.W. and T.G. developed the methods of analysis. S.W. performed all the calculations. All authors contributed equally to analyzing the results, writing and reviewing the paper.

\addcontentsline{toc}{section}{References}

\addcontentsline{toc}{section}{Appendix}
\begin{appendices}

\section{Relation between power law and scaling invariance}
\label{appa}
\setcounter{figure}{0}
\renewcommand{\thefigure}{\thesection\arabic{figure}}
\setcounter{equation}{0}
\renewcommand{\theequation}{\thesection\arabic{equation}}
\setcounter{table}{0}
\renewcommand{\thetable}{\thesection\arabic{table}}

For the convenience of the reader, we summarize salient features in Ref.~\cite{Newman2005}. We use the notation $\mathcal{R}(l)=\big \langle R_{ij}(\tau,l |l_{\omega})\big \rangle_{i}$ for each given $\tau$ within the maximal reachable distance $l_\omega$. If the response in terms of distances is scaling invariant, it fulfills the property
\begin{equation}
\mathcal{R}(l)=\mu(\lambda)\mathcal{R}(\lambda l) \ ,
\label{eqa1}
\end{equation}
where $\lambda$ is a scaling factor and $\mu(\lambda)$ is a function in terms of $\lambda$. Setting $l=1$ in the above equation gives
\begin{equation}
\mu(\lambda)=\frac{\mathcal{R}(1)}{\mathcal{R}(\lambda)}
\label{eqa2}
\end{equation}
Therefore Eq.~\eqref{eqa1} becomes
\begin{equation}
\mathcal{R}(\lambda)\mathcal{R}(l)=\mathcal{R}(1)\mathcal{R}(\lambda l) \ .
\label{eqa3}
\end{equation}
By differentiating both sides with regard to $\lambda$, we have
\begin{equation}
\mathcal{R}'(\lambda)\mathcal{R}(l)=l\mathcal{R}(1)\mathcal{R}'(\lambda l) \ .
\label{eqa4}
\end{equation}
Here $\mathcal{R}'(\cdot)$ represents the derivative of 
$\mathcal{R}(\cdot)$ regarding its argument inside the bracket. Letting $\lambda=1$ gives rise to
\begin{equation}
\mathcal{R}'(1)\mathcal{R}(l)=l\mathcal{R}(1)\mathcal{R}'(l)= l\mathcal{R}(1)\frac{d\mathcal{R}(l)}{dl}\ .
\label{eqa5}
\end{equation}
We rewrite Eq.~\eqref{eqa5} 
\begin{equation}
\frac{d\mathcal{R}(l)}{\mathcal{R}(l)}=\frac{\mathcal{R}'(1)}{\mathcal{R}(1)}\frac{dl}{l}\ .
\label{eqa6}
\end{equation}
Integrating both sides results in
\begin{equation}
\ln \mathcal{R}(l)=\frac{\mathcal{R}'(1)}{\mathcal{R}(1)} \ln l +c \ ,
\label{eqa7}
\end{equation}
where $c$ is a constant. Let l=1 such that we are able to obtain $c=\ln \mathcal{R}(1)$. This leads to
\begin{equation}
\mathcal{R}(l)=\mathcal{R}(1)l^\beta \ ,
\label{eqa8}
\end{equation}
where $\beta=\mathcal{R}'(1)/\mathcal{R}(1)$. Therefore, the scaling invariance results in the power-law response function in terms of distances. In other words, the power-law response function in terms of distances is the only function that meets the scaling invariant criterion~\eqref{eqa1}. 

\section{Figures for different time periods}

\setcounter{figure}{0}
\renewcommand{\thefigure}{\thesection\arabic{figure}}
\setcounter{equation}{0}
\renewcommand{\theequation}{\thesection\arabic{equation}}
\setcounter{table}{0}
\renewcommand{\thetable}{\thesection\arabic{table}}

\begin{figure}[htb]
\begin{center}
\includegraphics[width=\textwidth]{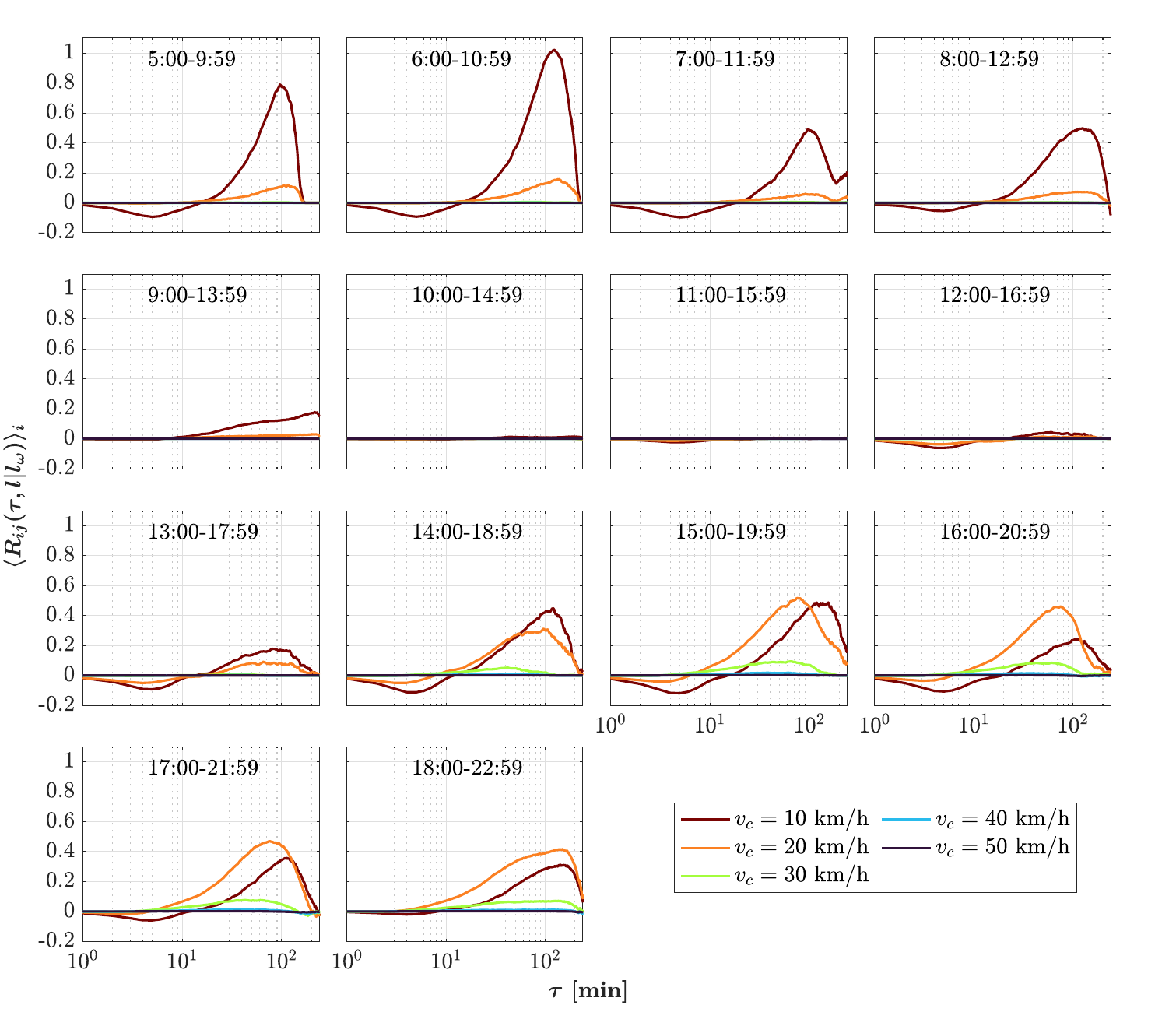}
\caption{Responses $\langle R_{ij}(\tau,l |l_{\omega})\rangle_{i}$ versus time lag $\tau$ within the distance range $l=2$ km under different critical velocities $v_c$ from 10 km/h to 50 km/h. }
\label{figb1}
\end{center}
\end{figure}

\begin{figure}[htb]
\begin{center}
\includegraphics[width=\textwidth]{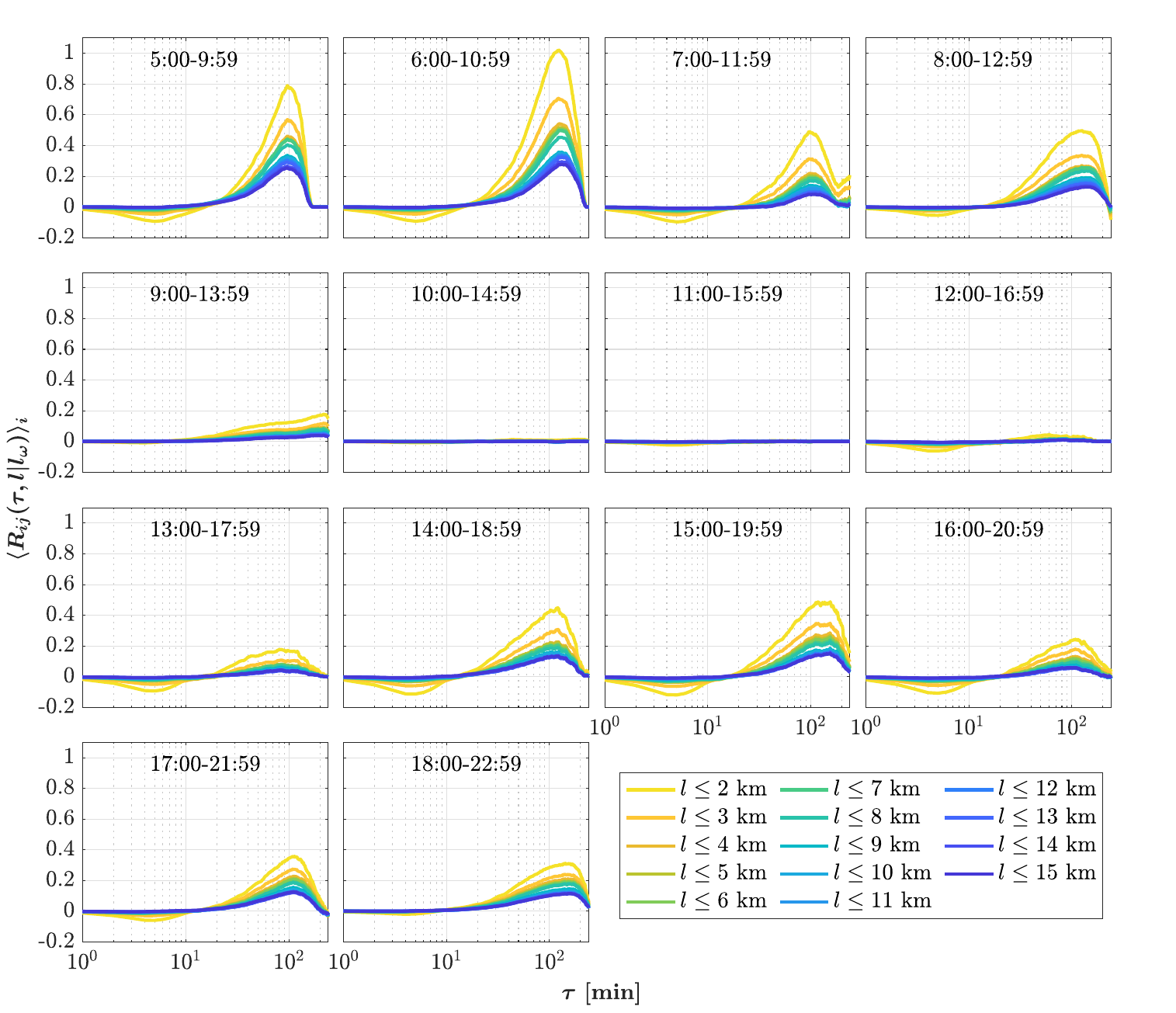}
\caption{Responses $\langle R_{ij}(\tau,l |l_{\omega})\rangle_{i}$ versus time lag $\tau$ within each distance range $l$ from 2 km to 15 km under the critical velocity $v_c=10$ km/h. }
\label{figb2}
\end{center}
\end{figure}

\begin{figure}[htb]
\begin{center}
\includegraphics[width=\textwidth]{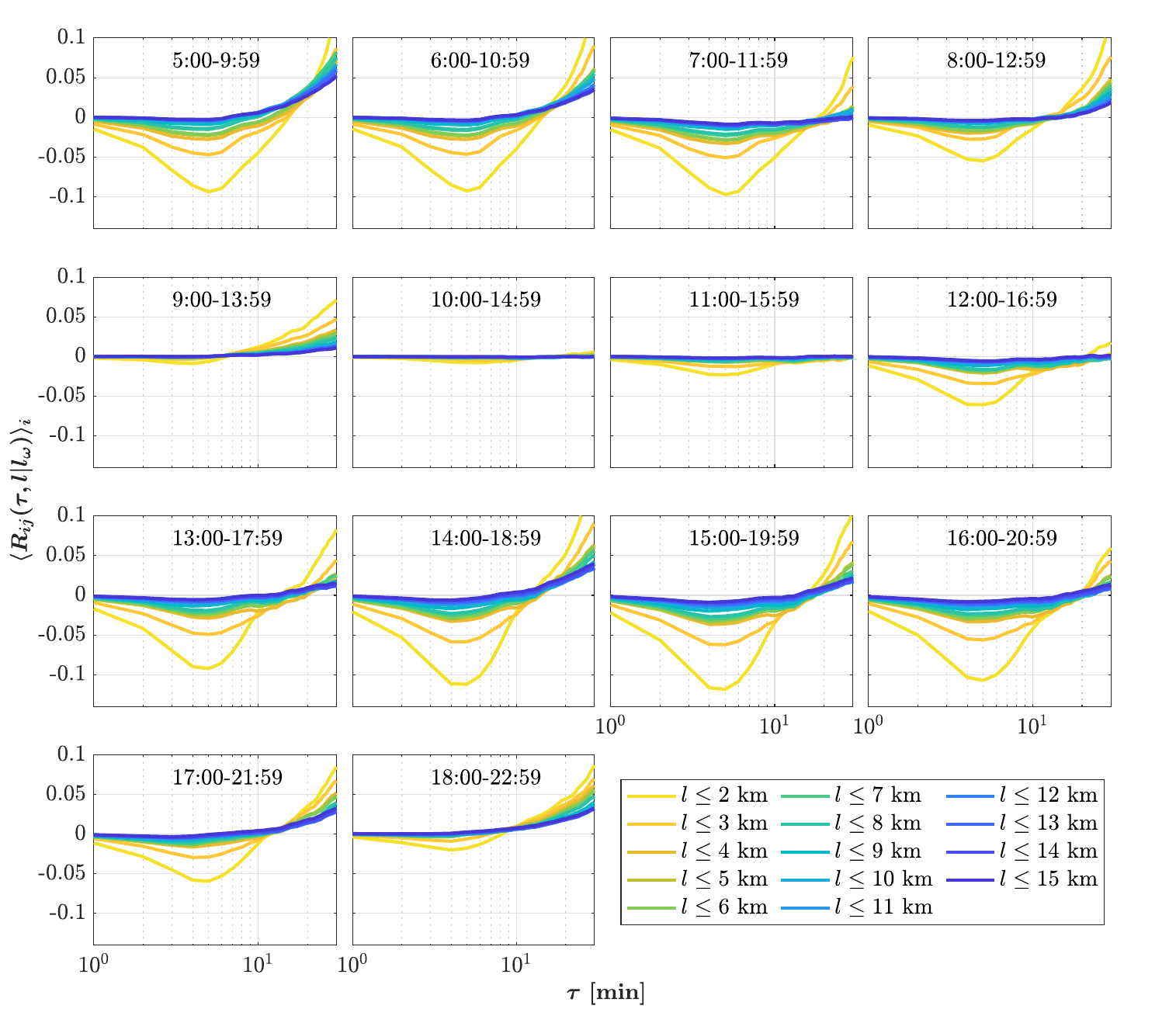}
\caption{The enlargement of negative responses $\langle R_{ij}(\tau,l |l_{\omega})\rangle_{i}$ versus time lag $\tau$ for $l$ from 2 km to 15 km and $v_c=10$ km/h.}
\label{figb3}
\end{center}
\end{figure}

\begin{figure}[htb]
\begin{center}
\includegraphics[width=\textwidth]{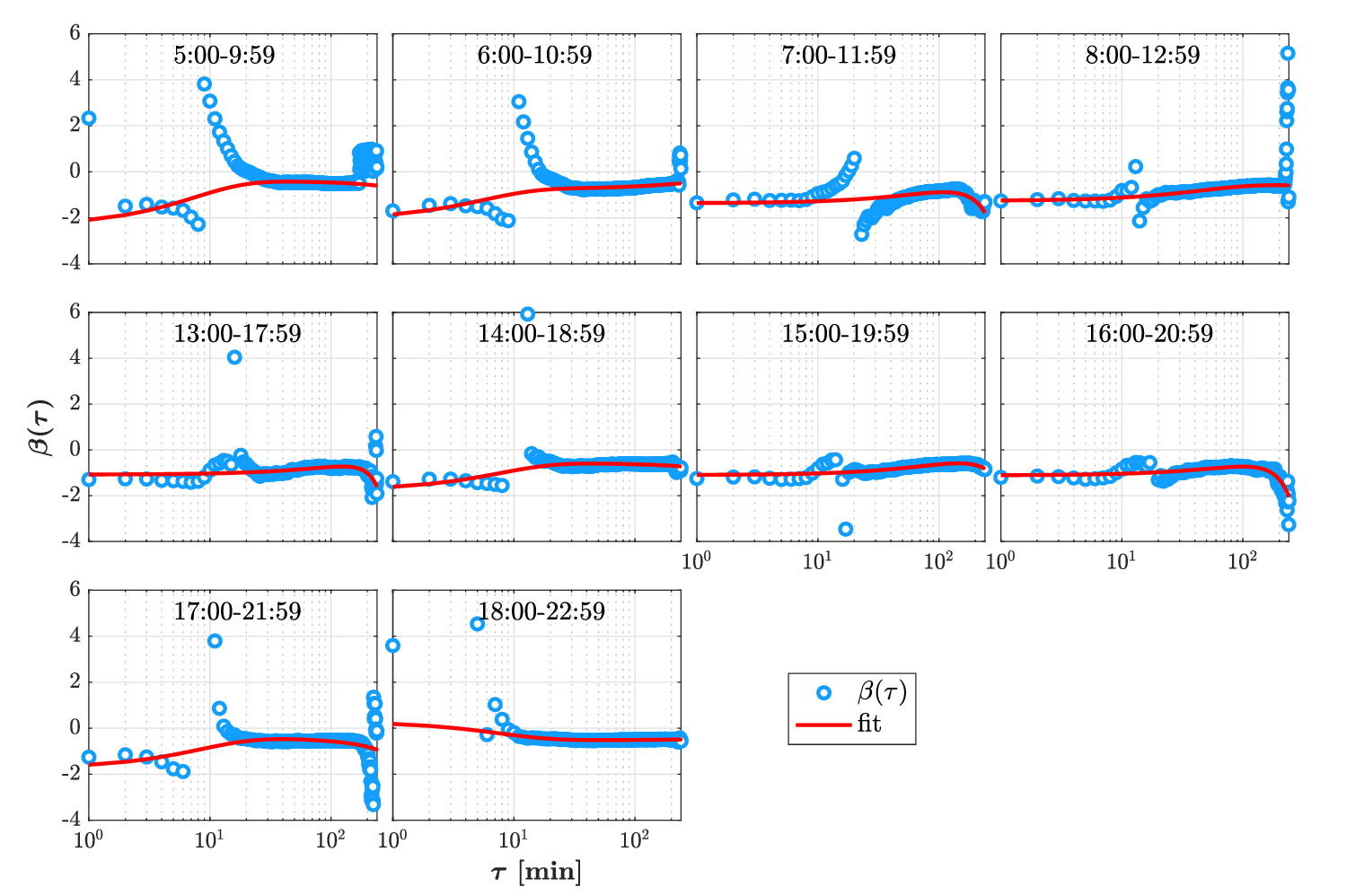}
\caption{Exponent $\beta(\tau |l_{\omega})$ versus time lag $\tau$ fitted by exponential function~\eqref{eq4.4.6}, where $v_c=10$ km/h.}
\label{figb4}
\end{center}
\end{figure}

\begin{figure}[tb]
\begin{center}
\includegraphics[width=\textwidth]{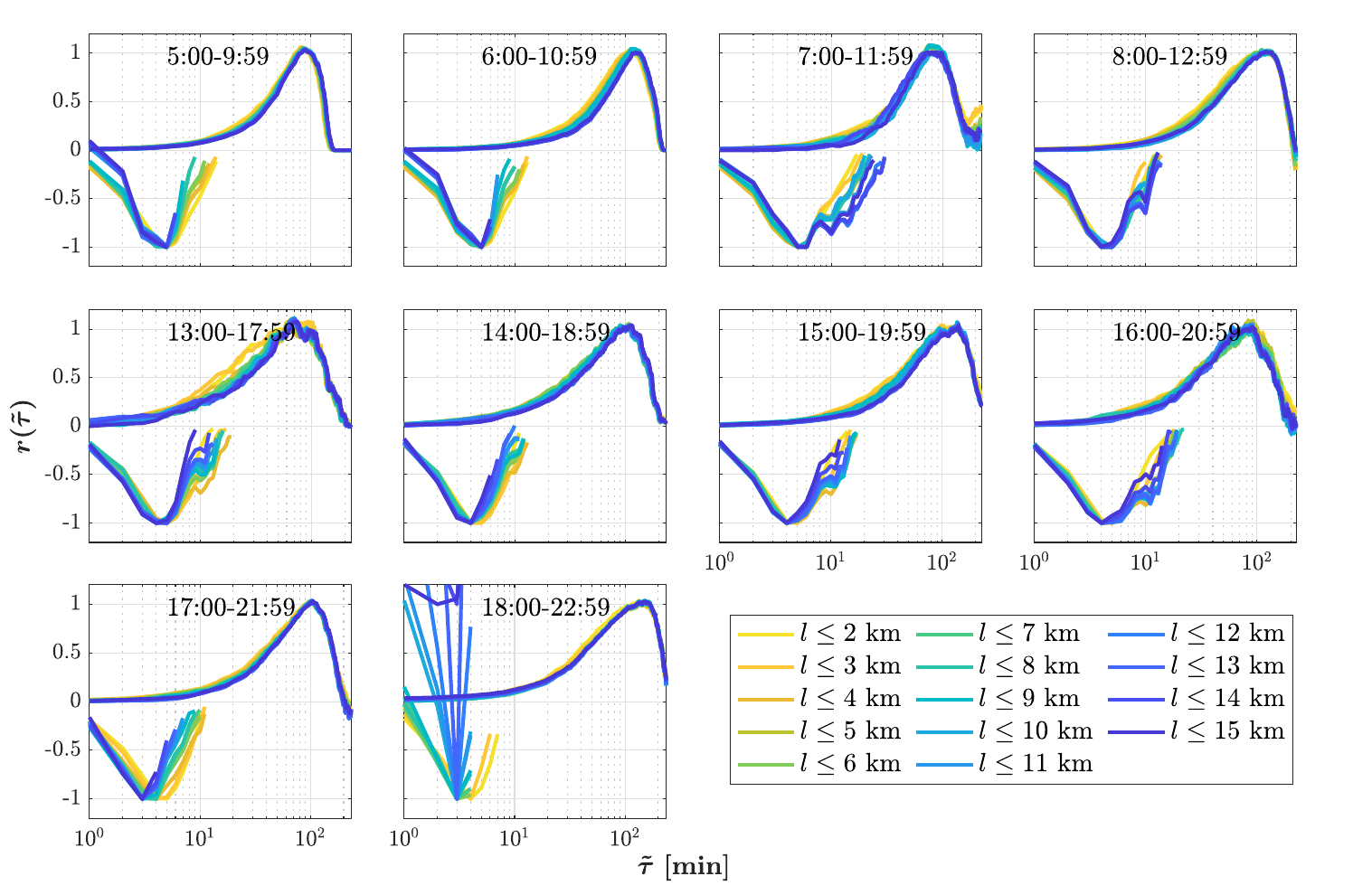}
\caption{Rescaled responses $r(\tilde{\tau})$ versus time lags $\tilde{\tau}$ within different distance ranges $l$ during 10 time periods, where $\tilde{\tau}=\tau$ and $r(\tilde{\tau})< 0$ for $\tau<\tau_c$, and  $\tilde{\tau}=\tau-\tau_c$ and $r(\tilde{\tau})\geq  0$ for $\tau\geq \tau_c$.}
\label{figb5}
\end{center}
\end{figure}

\begin{figure}[htb]
\begin{center}
\includegraphics[width=\textwidth]{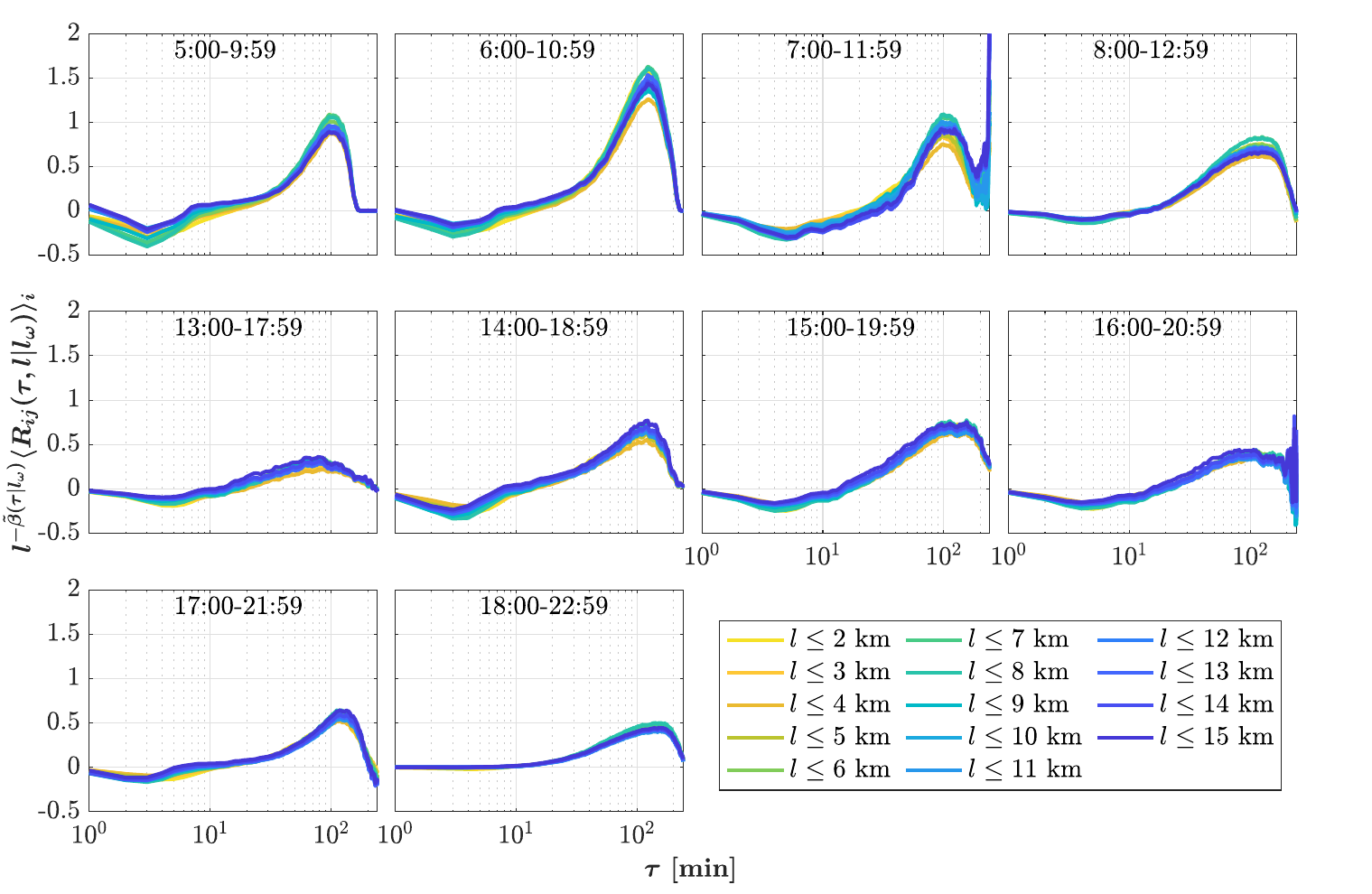}
\caption{Time evolution of rescaled responses $l^{-\tilde{\beta}(\tau |l_{\omega})} \langle R_{ij}(\tau,l |l_{\omega})\rangle_{i}$ within different distance ranges $l$ almost collapsing to the same curve during 10 time periods, where $\tilde{\beta}(\tau |l_{\omega})$ results from the fitting with the exponential function~\eqref{eq4.4.6} and $v_c=10$ km/h.}
\label{figb6}
\end{center}
\end{figure}

\end{appendices}

\end{document}